\documentstyle[mnextra,astrobib,graphics]{mn-ab}
\seceqnum           
\def\eg{{\rm e.g. }}
\def\ie{{\rm i.e. }}

\def\lsim{\mathrel{\hbox{\rlap{\hbox{\lower4pt\hbox{$\sim$}}}\hbox{$<$}}}}
\def\gsim{\mathrel{\hbox{\rlap{\hbox{\lower4pt\hbox{$\sim$}}}\hbox{$>$}}}}

\title[Constraining $\Omega_{0}$ using weak
gravitational lensing by clusters]
      {Constraining $\Omega_{0}$ using weak
gravitational lensing by clusters}
\author[G. Wilson, S. Cole, and C. S. Frenk] 
       {Gillian Wilson$^{1}$, Shaun Cole$^{2}$, Carlos S. Frenk$^{3}$ \\ 
Department of Physics, University of Durham, Science
Laboratories, South Rd, Durham DH1 3LE \\
$^1$Gillian.Wilson@durham.ac.uk 
$^2$Shaun.Cole@durham.ac.uk 
$^3$C.S.Frenk@durham.ac.uk
}
  
\begin{document}

\maketitle

\begin{abstract}

\noindent The morphology of galaxy clusters reflects the epoch at which
they formed and hence depends on the value of the mean cosmological
density, $\Omega_0$. Recent studies have shown that the distribution of
dark matter in clusters can be mapped from analysis of the small
distortions in the shapes of background galaxies induced by weak
gravitational lensing in the cluster potential. We construct new statistics
to quantify the morphology of clusters which are insensitive to 
limitations in the mass reconstruction procedure. By simulating weak
gravitational lensing in artificial clusters grown in numerical simulations
of the formation of clusters in three different cosmologies, we obtain
distributions of a quadrupole statistic which measures global
deviations from spherical symmetry in a cluster. These distributions are
very sensitive to the value of $\Omega_0$ and, as a result, lensing
observations of a small number of clusters should be sufficient to place
broad constraints on $\Omega_{0}$ and certainly to distinguish between the
extreme values of 0.2 and 1.

\end{abstract}


\section{Introduction}

In a low density universe, density fluctuations cease to grow after a
redshift $z \sim\frac{1}{\Omega_{0}}-1$, where $\Omega_0$ is the present
value of the cosmological density parameter (see \eg \citeNP[\S11 \&
\S13]{peeb-80}). Introducing a cosmological constant, $\lambda_0$ (which we
shall express in units of $3H_0^2$, where $H_0$ is the present value of the
Hubble constant) makes this cessation more abrupt.  Hence, in low
$\Omega_{0}$ models, both with and without a cosmological constant,
clusters form at moderate redshift ($z$ = 1--4), and subsequently accrete
very little material. In this period, which can span many dynamical times,
the internal structure of the clusters relaxes to produce smooth, nearly
spherical configurations. In contrast, in $\Omega_{0}=1$ models structure
formation occurs continuously, rich galaxy clusters form in abundance only
at very low redshift ($z$ = 0.2--0.3), and continue to accrete material
even at the present epoch.  Hence, if $\Omega_{0}=1$, many clusters are
expected to show evidence of recent merger events and to have irregular
morphologies.

Over the past few years a number of optical and X-ray studies have
suggested that a significant proportion of clusters, perhaps $\sim 40 \%$,
show evidence of substructure (\eg
\citeNP{gel-82,dress-88,west-90,forman-90,bird-94,west-95}.
Motivated by these interesting but controversial observations suggesting
recent cluster growth, \shortciteN{evr-94} and \citeN{mohr-95}
demonstrated that, in agreement with analytic predictions
\cite{rich-92,lacey-93}, the internal structure of clusters is
sensitive to the cosmological model. \shortciteN{evr-94} constructed X-ray
surface brightness maps from their hydrodynamic N-body simulations of
cluster formation and compared these to X-ray maps of 65 clusters
observed with the Einstein Imaging Proportional Counter.  They concluded
that galaxy clusters with the observed range of X-ray morphologies were
more likely to have arisen in a high $\Omega_{0}$ cosmology.

In parallel with these developments, it has become clear that the structure
of the dynamically dominant dark matter in clusters can be reliably mapped
by analysing the distortions in the images of background galaxies induced
by weak gravitational lensing in the cluster potential. Our aim in this
paper is to assess whether the surface overdensity maps revealed by such
analyses may be used to quantify cluster morphology and hence to constrain
$\Omega_{0}$. To this end we simulate the lensing of background galaxies
by clusters drawn from three different cosmological models. These are the
same clusters studied by \shortciteN{evr-94}.  For each cluster we create
a mock CCD frame as described in \citeN{wil-96a} (hereafter WCF) and then
use the inversion technique of \citeN{ks-93} to reconstruct 
a map of the projected cluster overdensity.

The remainder of this paper is organised as follows. The main features of
the Kaiser \& Squires reconstruction method are summarized in
Section~\ref{ssec:lens rec}.  Section~\ref{ssec:shapes} introduces new
statistics for measuring cluster shapes, specifically tailored to be
insensitive to uncertainties inherent in the reconstructed
surface overdensity maps. In Section~\ref{ssec:clus sim} we describe the
simulated galaxy clusters to be used as lenses and whose density profiles
we examine. Section~\ref{ssec:CCD} reviews how we generate the background
distribution of galaxies and create mock CCD frames of the lensed
galaxies. Our results are presented in Section~\ref{sec:results} where we
compare the reconstructions to the original clusters and assess how the
measured distribution of cluster shapes depends on the cosmological model.
We conclude in Section~\ref{sec:end} with a discussion and summary of our 
main results.

\section{Methods}
\label{sec:methods}

\subsection{Lensing Reconstruction}
\label{ssec:lens rec}

The lensing reconstruction technique employed in this paper was devised by
\citeN{ks-93} (hereafter KS). It produces a map of the estimated
overdensity, $\hat{\sigma}(\vec{\theta})$, where
$\hat{\sigma}(\vec{\theta})$ is the deviation of the true cluster surface
density, $\hat{S}$, from the mean surface density, $\bar S$, within the
area being considered, measured in units of the critical surface density
$S_{\rm crit}$:
\begin{equation}
\label{eq:sdest}
\hat{\sigma}(\vec{\theta})=\frac{\hat{S}-
 {\bar S}}{S_{\rm{crit}}}, 
\end{equation}
where $\vec{\theta}$ is the position vector on the plane of the lens. 
\noindent{$S_{\rm crit}$ is
defined as}
\begin{equation}
\label{eq:scrit}
 S_{\rm{crit}} = \frac{c^{2}}{4 \pi G } \frac {D_{\rm{os}}}
{D_{\rm{ol}}D_{\rm{ls}}}, 
\end{equation}
where $D$ denotes angular-diameter distance and the subscripts refer to the
observer, lens and source galaxy. This critical surface density is the
minimum required to produce multiple images of a source object.

As shown in WCF, the KS technique recovers the mass 
distribution in clusters extremely well when the lensing is weak \ie when
the bending angle varies slowly with position. Our CCD simulations in that
paper demonstrated that observational complications such as seeing and
noise act to diminish the reconstructed surface density.  However, we
showed that it was possible to correct fully for this diminution and
proposed a method to estimate a multiplicative compensation factor, f.

\subsection{Quantifying Shapes}
\label{ssec:shapes}

Lensing reconstructions such as that of KS reliably return {\it relative}
values of the surface overdensity, $\hat{\sigma}(\vec{\theta})$
(equation~\ref{eq:sdest}), but do not fix its {\it absolute} value. 
Since $S_{\rm crit}$ depends on
the geometry of the lensing configuration through the angular-diameter
distance relationship, the redshift distribution of source galaxies must be
known before the mean value of $S_{\rm crit}$
(equation~\ref{eq:scritbar} below) and hence the {\em absolute} surface
overdensity $\hat{S}- \bar{S}$ can be obtained from
equation~\ref{eq:sdest}.  In addition, the KS technique is only sensitive
to variations in surface density from the mean. This is because a uniform
slab of material across the whole lens plane does not distort the images
of galaxies lying behind. Thus, the mean surface density, $\bar S$, is
also unknown unless the region analysed is sufficiently large to encompass
the whole of the lensing material so that the surface density near the
edge of the region can be taken as the zero-point.

In order to minimise the effects of these uncertainties, we define dipole
and quadrupole statistics which depend not on the {\em absolute} surface
overdensity, but only on the values of $\hat{\sigma}(\vec{\theta})$
relative to each other. These statistics, therefore, are explicitly
independent of $\bar S$, $S_{\rm crit}$ and any compensation factor $f$. We
define the dipole, $D(A)$, by
\begin{equation}
D(A) = \frac{[d_{1}^{2}+d_{2}^{2}]^{\frac{1}{2}}}{A^{\frac{3}{2}}}  
= \frac{\left| d \right|}{A^{\frac{3}{2}}} 
\end{equation}
\noindent{where}
\begin{equation}
d_{1}  = \int H(\hat{\sigma}-\sigma_{\rm con})x \,dA ,
\end{equation}
\begin{equation}
d_{2}  = \int H(\hat{\sigma}-\sigma_{\rm con})y \,dA 
\end{equation}
and
\begin{equation}
A  = \int H(\hat{\sigma}-\sigma_{\rm con}) \,dA .
\end{equation}
Here, $\sigma_{\rm con}$ is the surface overdensity of the contour
level above which we choose to evaluate the statistic and $A$ is the
area within that contour. $H$ is the Heaviside step function,
$H(\hat{\sigma}-\sigma_{con})$, which is equal to 0 if $\hat{\sigma} <
\sigma_{\rm con}$ and equal to 1 if $\hat{\sigma} \geq \sigma_{\rm
con}$. The integrals are evaluated over the area within the contour. 
Similarly the quadrupole, $Q(A)$, for the
same area $A$ is defined as
\begin{equation}
Q(A) = \frac{[q_{1}^{2}+q_{2}^{2}]^{\frac{1}{2}}}{A^{2}}  
= \frac{\left| q \right|}{A^{2}} 
\end{equation}
\noindent{where}
\begin{equation}
q_{1} = \int H(\hat{\sigma}-\sigma_{\rm con})(x^{2}-y^{2}) \,dA 
\end{equation}
and
\begin{equation}
q_{2} = \int H(\hat{\sigma}-\sigma_{\rm con})2xy \,dA  .
\end{equation}
The $x$ and $y$ coordinates are measured relative to the cluster centre.
(We shall discuss exactly how we define the cluster centre in
Section~\ref{sec:results}.) In practice, the integrals become sums over
pixels in the reconstructed maps. Note that both these statistics 
are dimensionless. If one pictures a reconstructed surface density map as
a contour plot, then these statistics are independent of the value of
$\hat\sigma$ labelling each contour because we select the contour by the
area $A$ that it encloses rather than by its level. Thus, once a suitable
area $A$ has been chosen, say 5 arcmin$^{2}$, the statistics are
independent of :- \begin{enumerate}
\item Any dilution of the lensing signal due to seeing.
\item Nonlinearity effects which may result in an underestimate of 
the surface overdensity in regions of high $\hat{S}$. 
\item The (unknown) value of $S_{\rm crit}$.
\end{enumerate}
The values of the statistics $D(A)$ and $Q(A)$ will, of course, 
depend on the noise level in the reconstructed surface
density maps. It is this effect on these statistics that we aim to quantify
with our simulations of lensing.

\subsection{The Cluster Simulations}

\label{ssec:clus sim}

As our lenses we used a set of eight N-body gasdynamic simulations
of the formation of galaxy clusters, further details of which may be
found in \shortciteN{evr-94}.  The clusters evolved from the same
eight sets of initial density fields but in three different
cosmologies: 

\begin{enumerate}
\item A biased Einstein-de Sitter model
[ $\Omega_{0}=1$, $\sigma_{8}=0.59$, where $\sigma_{8}$
is the rms fluctuation of mass within spheres of radius 
$8h^{-1}{\rm Mpc}$, and $h$ is the present value of the
Hubble constant in units of $100$ kms$^{-1}$Mpc$^{-1}$].

\item An unbiased ($\sigma_{8}=1.0$) open model with $\Omega_{0}=0.2$
and $\lambda_{0}=0$.

\item An unbiased ($\sigma_{8}=1.0$) low density flat model with
 $\Omega_{0}=0.2$ and $\lambda_{0}=0.8$.
\end{enumerate}

In each of four periodic boxes of size $L=15,20,25$ and $30$ $h^{-1}$ Mpc, 
\shortciteANP{evr-94} created two constrained realizations of Gaussian
random density fields, with the standard cold dark matter power
spectrum appropriate for $H_{0} =50$ km ${\rm s^{-1}}$ ${\rm
Mpc^{-1}}$.  In each case, using the technique of \citeN{bert-87},
they imposed the constraint that, when smoothed with a Gaussian of width 
$R_f= 0.2L$, there be a peak at the centre of the box with height
2.5--5 times the {\it rms} density fluctuation on this scale.

Evrard's (1988) P$^{3}$M/SPH hydrodynamic N-body code was used to evolve
the particle distributions to the present epoch.  Two sets of $32^{3}$
particles represented the dark matter and gas respectively. A baryon
content of $\Omega_{\rm b}=0.1$ was assumed for all the models. Gravity,
PdV work and shock heating were incorporated for the gas but the effects
of radiative cooling were ignored. The spatial resolution was
approximately $0.005L$, varying from 75 to 150 $h^{-1}$ kpc depending on
the box length. For our lensing simulations we chose the output epoch from
each cosmology closest to redshift $z=0.18$. This value is fairly typical
in observational studies \cite{wil-96c} and corresponds to the value
adopted in our previous simulations (WCF).

In the simulations of \shortciteN{evr-94} the 
mass of each cluster was approximately proportional to the volume of the
simulation box. Our aim here is to study the morphology of clusters
detected by gravitational lensing with similar signal-to-noise ratio in
each cosmology. Thus, for our purposes it is more convenient to have a set
of clusters all of the same mass. To achieve this we simply rescaled the
mass of each particle in the smaller clusters so that the total mass was
the same as that in the largest cluster \ie the one in the
$L=30 h^{-1}$ Mpc box. This required multiplying each mass by the factor $(30 
h^{-1} {\rm Mpc}/L)^{3}$. In order to preserve the correct density, we also
multiplied the simulation length by $30 h^{-1} {\rm Mpc}/L$.

The typical mass of a rich cluster is known approximately from
observations. For a fair intercomparison of the models, the clusters
considered should all have approximately the same mass within a fiducial
radius such as the Abell radius ($1.5 h^{-1}$Mpc). The simulated clusters
in the open and flat $\Omega_{0}=0.2$ cosmologies were somewhat less
massive and so we scaled them as in the previous paragraph to have
approximately the same total mass as the average $\Omega_{0}=1$ cluster
($\sim 1.36 \times 10^{15} h^{-1} {\rm M_\odot}$ within an Abell
radius). The maximum scaling required was a factor of 2.4 in mass.
 
After this rescaling the initial conditions are no longer appropriate for a
standard CDM model but instead have power spectra with slopes which differ
slightly from the correct ones. However, analytic arguments show that
moderate changes in the slope of the power spectrum are far less important
in determining the formation history of a cluster than the actual value of
$\Omega_0$ (Lacey \& Cole 1993). Furthermore, recent high-resolution
simulations of the formation of dark matter halos in an $\Omega_0=1$
universe by Cole \& Lacey (1995) demonstrate that a statistic similar to
the one we use below to characterize cluster shapes is insensitive to
spectral index and halo mass (cf their Figure~17). Thus, our rescaling
should have a negligible effect on our results.

We tripled the size of our cluster sample by choosing three perpendicular
axes through the centre of each cluster and treating the projection along
each axis as a different cluster. This resulted in 72 clusters, 
24 from each of the three cosmologies.

\begin{figure*}
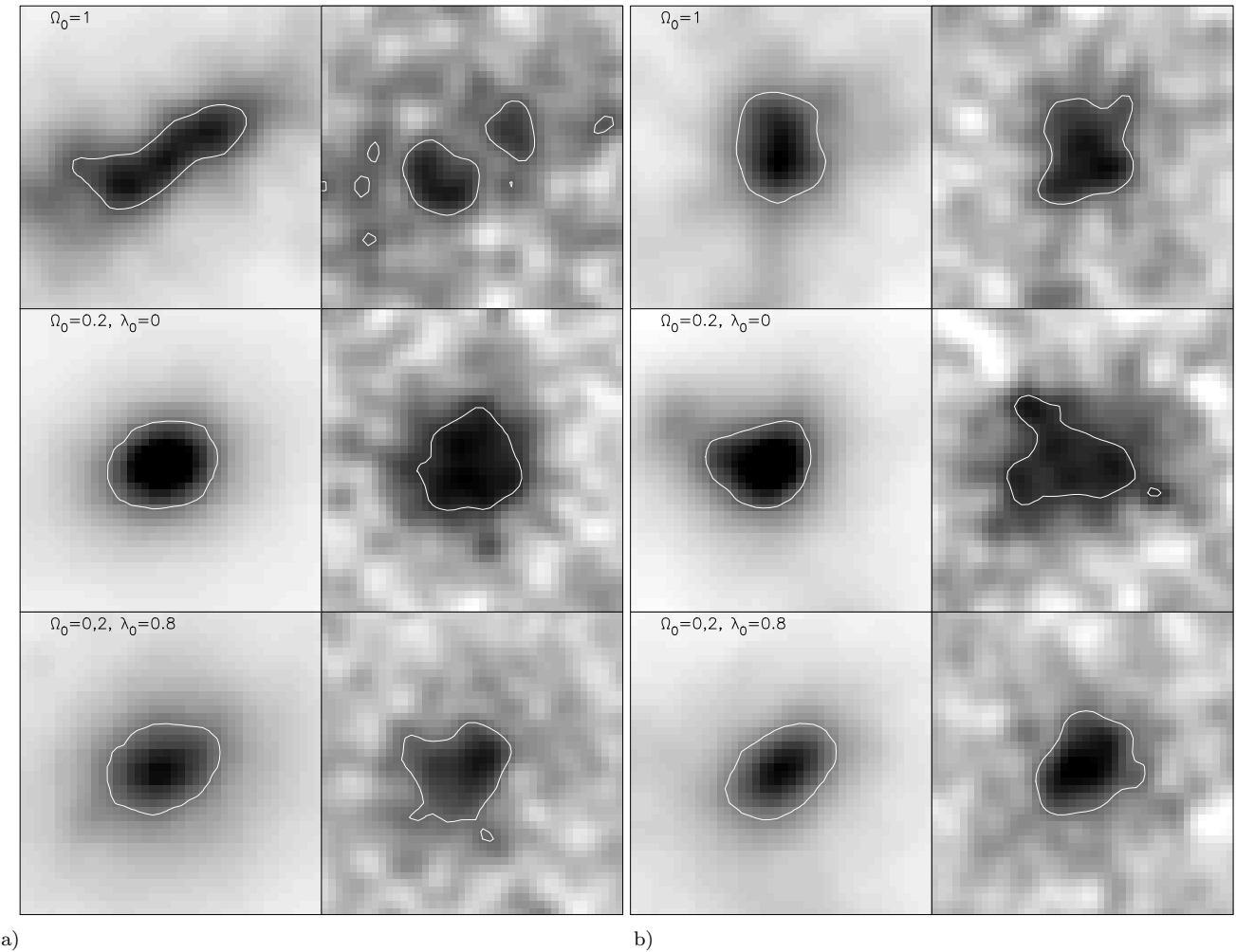

\begin{center}
\scalebox{0.49}{
 \includegraphics[25,35][517,780]{wcf2figs/figure1.eps}
 \includegraphics[25,35][517,780]{wcf2figs/figure2.eps}
}
\leftline{ a) \hskip8.5cm  b) }
\caption{Examples of surface overdensity maps for four clusters from each
of the three cosmologies. In each group the left-hand column shows the
original surface density maps and the right-hand column the corresponding
reconstruction. The greyscales have been adjusted according to the central
surface overdensity in each case. From top to bottom the clusters are from
the $\Omega_0=1$, open $\Omega_0=0.2$ and flat $\Omega_0=0.2$ models with
the same initial fluctuation spectrum. The original CCD frame from which the
reconstruction was made subtended an angle of $\sim$ 10 arcmins but these
maps show only the inner 6.7 by 6.7 arcmins. At the cluster redshift,
$z=0.18$, 1 arcmin is equivalent to $0.12 h^{-1}$Mpc for the $\Omega_0=1$
cosmology. The contour plotted encloses an area of 3.8 arcmin$^2$. A 
Gaussian smoothing of width $\theta_{\rm sm}$ = 0.25 arcminutes has been
used in the KS reconstruction procedure.
}
\label{fig:figure1}
\end{center}
\end{figure*}

\subsection{CCD images}
\label{ssec:CCD}

To simulate typical observational datasets, we constructed artificial
B-band CCD images of lensed field galaxies, building them up pixel by
pixel, in the manner described in WCF.  We employed the same distributions
of galaxy ellipticity, scalelength, redshift, magnitude, noise and seeing
as in that paper:

\noindent $\bullet$ Redshift and magnitude distributions

For the redshift distribution of the model galaxies, we adopted the $\rm
{m_{B}}=25$ distribution predicted by the analytic model of galaxy
formation of \citeN{cole-94}. Since the critical density, $S_{\rm crit}$,
depends on the source redshifts through equation~(\ref{eq:scrit}), we 
sampled this redshift distribution discretely and produced a set of source
planes spanning a range of redshifts. The net effect on the
reconstruction equations is that $S_{\rm crit}$ is replaced by the mean value,
${\bar S}_{\rm crit}$, defined by
\begin{equation}
\label{eq:scritbar}
{\bar S}_{\rm crit}^{-1}= \int \frac{1}{S_{\rm crit}(z)} \,p(z) \,dz ,
\end{equation}
where $p(z)$ is the probability that a galaxy lies at redshift $z$. For our
adopted redshift distribution, $\bar S_{\rm crit} \simeq 6 \times 10^{15} h
{\rm M_\odot / Mpc^{2}}$, with the exact value depending on the cosmology.
The distribution of apparent magnitudes we generated directly from the
B-band source counts of \citeN{met-95}.

\noindent $\bullet$ Scalelength distribution 

We assumed that all the background galaxies are disks with
exponential profiles.  The scalelength was chosen from a uniform
distribution in the range 0.25~arcseconds to 0.65~arcseconds, as
suggested by the observations of \citeN{tyson-94}.

\noindent $\bullet$ Ellipticity distribution 

Ellipticities for the background galaxies were generated by randomly
sampling the empirical ellipticity distribution derived from a
single frame in 0.7--0.9~arcseconds seeing by \citeN{brain-95}.

We created one CCD frame per cluster projection and then analysed it using
FOCAS \cite{jarvis-81} as described in WCF.  We assumed moderately good
seeing of 1~arcsecond full-width half-maximum.

\section{Results}
\label{sec:results}

\begin{figure*}
\begin{center}
\scalebox{0.49}{
 \includegraphics[25,35][517,780]{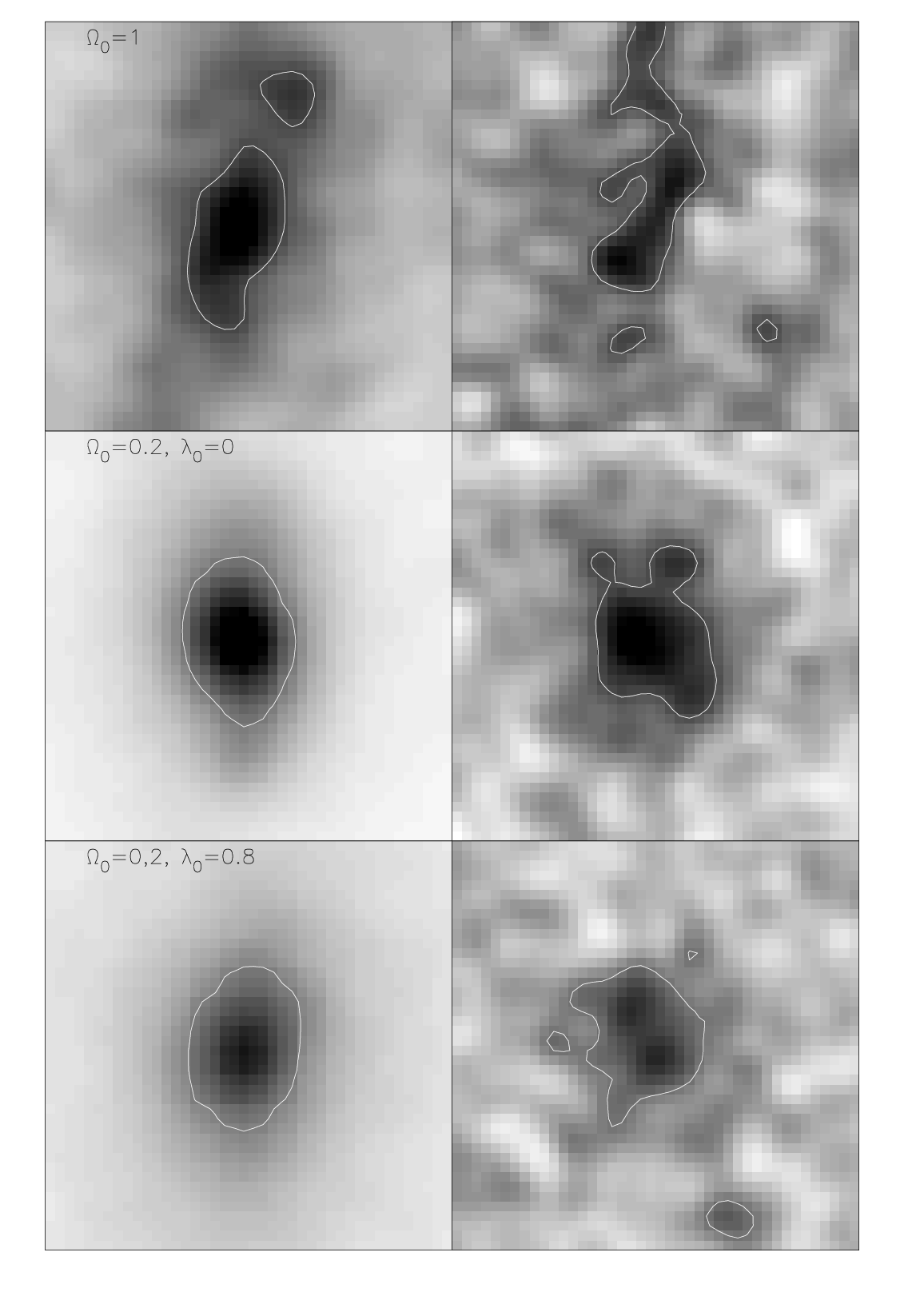}
 \includegraphics[25,35][517,780]{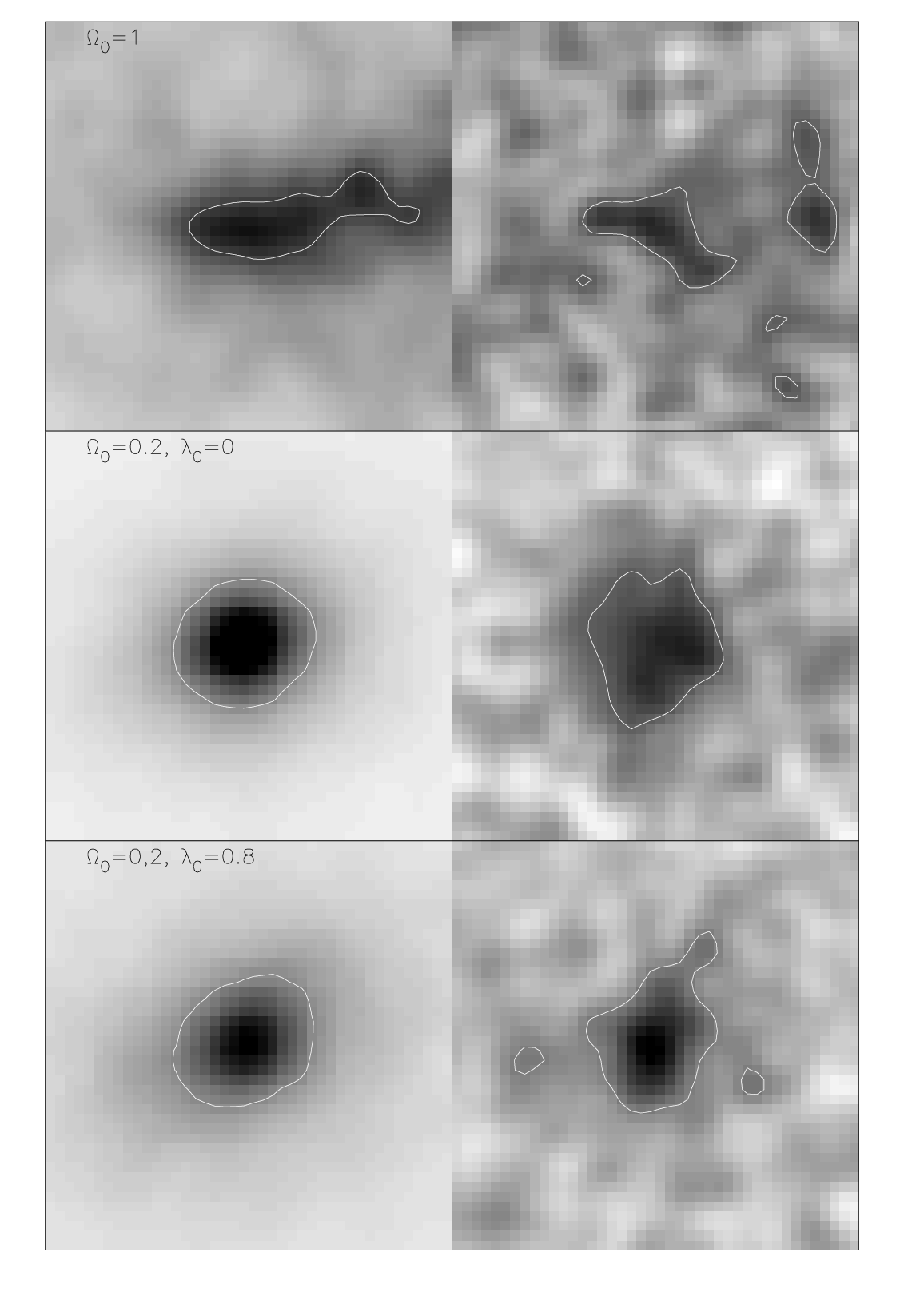}}
\leftline{ c) \hskip8.5cm  d) }
\leftline{ {\bf Figure~1.} continued.}
\end{center}
\end{figure*}

Figures~\ref{fig:figure1}a-d show examples of the projected mass
distribution in the central regions of some of our clusters. In each group,
from top to bottom, we illustrate clusters grown from identical  initial
conditions in the $\Omega_0=1$,
open $\Omega_0=0.2$, and flat $\Omega_0=0.2$ cosmologies respectively.  The
left-hand panels give the original surface density, smoothed with a
Gaussian of $\theta_{\rm sm}$ = 0.25 arcminutes, the same smoothing as we
employed in the KS reconstruction. The right-hand panels show the
corresponding reconstructions.  The field of view in each plot is
$6.7$~arcmins (equivalent to $0.8h^{-1}$ Mpc, for $\Omega_0=1$), and the
contour on each map encloses an area of 3.8 arcmin$^{2}$.

The surface overdensities on the left-hand side have been normalised so that
the mean value is zero over the full CCD frame, reflecting the fact that
the mean value from the KS reconstruction is automatically set to zero. To
enhance the substructure in the reconstructed maps, we have used different
greyscales in the left- and right-hand panels. The former appear
uniformly black near the centre. Limitations of scale, particularly in 
the open $\Omega_0=0.2$ case, conceal the reality
that the density distribution is in fact very highly peaked. The maximum
surface density can be critical or even greater. In contrast, the maximum
surface overdensity in the right-hand panels is $\simeq$ 0.2 or 0.3 of
critical. This diminution is partly due to observational effects such as
seeing and noise, and partly due to the effects of nonlinearity - weak
lensing reconstruction techniques generally underestimate $\hat{\sigma}$ in
regions where the surface overdensity rises above a few tenths of
critical. This failure of the method results in reconstructed profiles that
saturate near the cluster centre producing a broad plateau which, in
the open $\Omega_0=0.2$ model, extends over many
hundreds of pixels with similar density values. 

\begin{figure*}
\begin{center}
\scalebox{0.45}{
  \includegraphics[50,0][600,650]{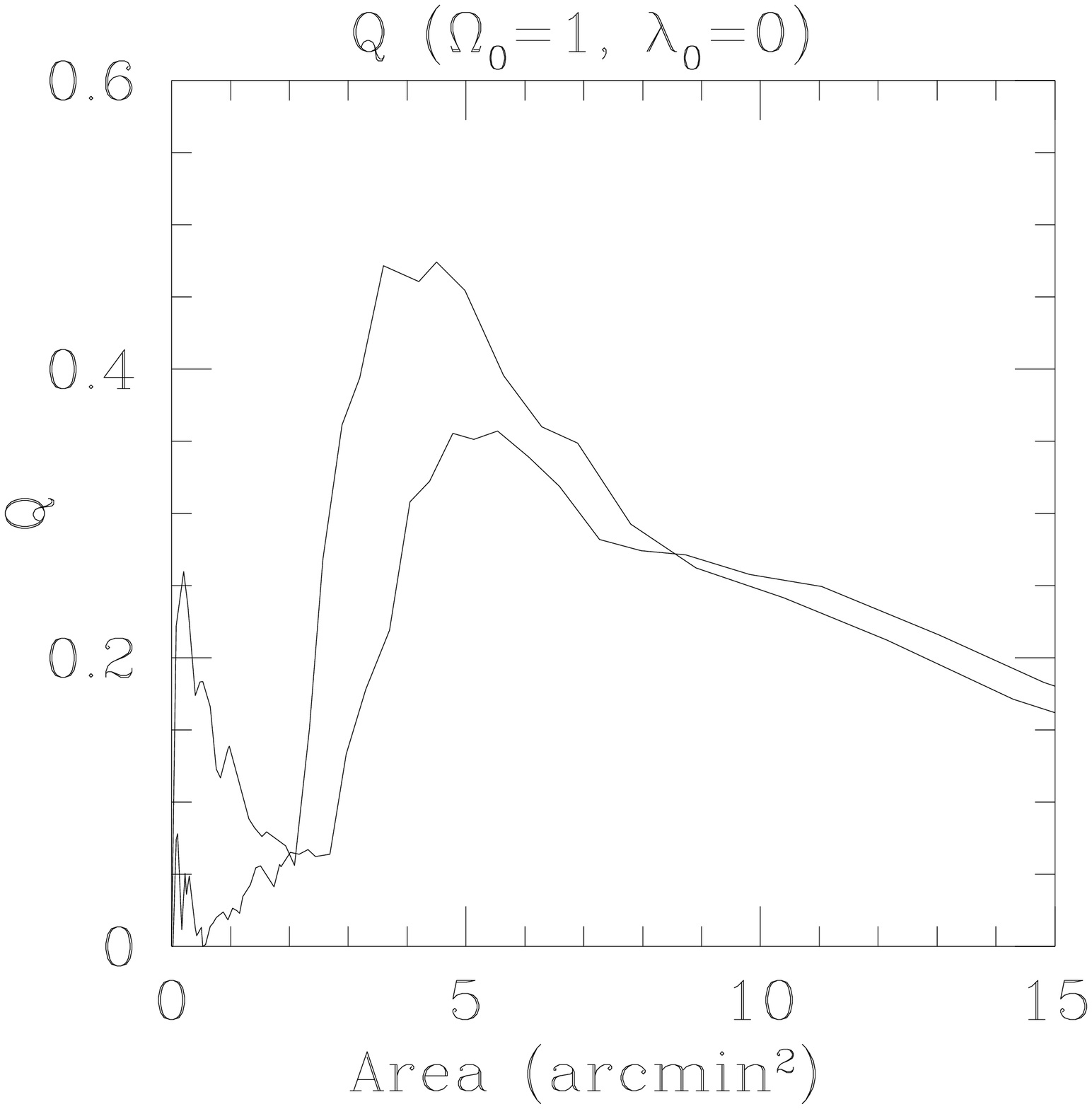}
  \includegraphics[50,0][600,650]{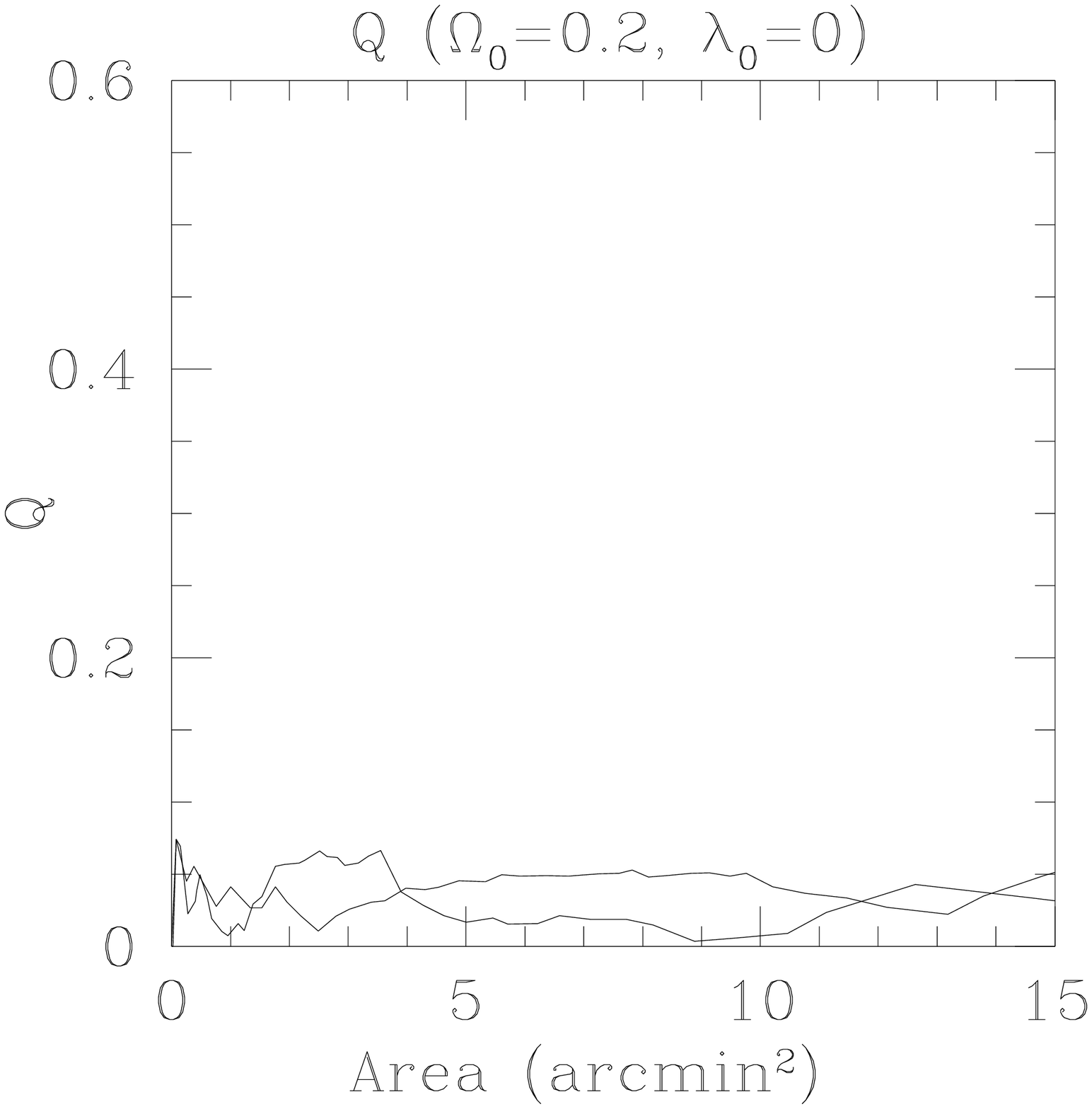}}
\leftline{ a) \hskip8.5cm  b) }
\caption{The quadrupole, $Q(A)$, as a function of the area over which it is
calculated. Two examples are shown of clusters in each of the $\Omega_0=1$
(left panel) and open $\Omega_0=0.2$ models (right panel).}
\label{fig:dipEdS}
\end{center}
\end{figure*}

Figure~1 illustrates how clusters in the $\Omega_0 = 1$ cosmology
are still forming at $z=0.18$. In most cases, large clumps are still
infalling and the dominant mass condensation is usually quite elongated.
By contrast, clusters in the open $\Omega_0 = 0.2$ cosmology are more
centrally concentrated and closer to spherical symmetry.  Clusters from the
flat $\Omega_0 = 0.2$ ($\lambda_0 = 0.8$) cosmology are similar to these
but they have slightly lower central overdensity and are a little more
elongated.  In general, the reconstructed shapes match the originals rather
well. With the exception of a few noisy pixels, the essential features of
the different cluster morphologies are preserved by the reconstruction.

We now use the dipole and quadrupole statistics, defined in
Section~\ref{ssec:shapes}, to quantify the distribution of cluster shapes
in each cosmology.  Initially we evaluated the statistics $D(A)$ and $Q(A)$
(equations~2.3 and 2.7) after identifying the cluster centre with the pixel
of maximum surface overdensity. We found that unless the reconstructions
have very high signal-to-noise, this choice of centre results in noisy
estimates of $D(A)$ and to a lesser extent of $Q(A)$. For clusters with
steep density profiles the reconstructed profiles have a saturated core
and, as a result, the highest density peak is simply the highest {\em
noise} peak anywhere in this central plateau. In light of this we tried
a second choice of cluster centre which is more robust than simply
selecting the densest pixel. This time we took the centre of coordinates to
be the position at which $D(A)=0$ (\ie the centroid of the area within the
contour). The choice of centre now depends on the area, $A$, selected, but
is much less sensitive to the presence of noise. Having chosen the centre
such that $D(A)=0$ we then use only $Q(A)$ as a measure of the asymmetry of
the cluster.

\begin{figure}
\begin{center}
\scalebox{0.45}
{\includegraphics[100,0][450,740]{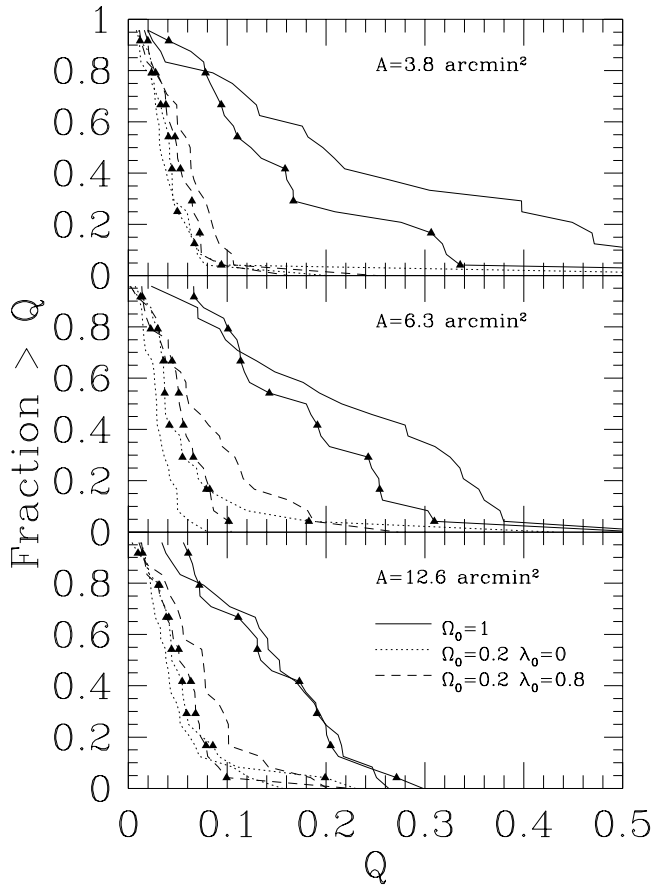}}
\caption{The cumulative probability distribution of the quadrupole, $Q(A)$,
at a fixed area, $A$, for $A=3.8$ arcmin$^2$ (top), $6.3$ arcmin$^2$
(middle), and $12.6$ arcmin$^2$ (bottom).  The cluster centre has been
chosen as the point about which the dipole vanishes.  The two solid lines
are for the $\Omega_0=1$ model, the two dotted lines for the open
$\Omega_0=0.2$ model, and the two dashed lines for the flat $\Omega_0=0.2$
model. In each case the curves marked by triangles correspond to the
original clusters while the unmarked lines correspond to the reconstructed
clusters. }
\label{fig:150}
\end{center}
\end{figure}

Figure~\ref{fig:dipEdS}a shows $Q(A)$ as a function of area, $A$, for two
clusters from the $\Omega_0=1$ cosmology. For $A<2$ arcmin$^{2}$, $Q(A)$
tends to be small because of the dominant contribution from the central
relaxed core, although its value fluctuates due to noise.  As larger
areas (and hence smaller surface overdensities) are considered, some of the
more distant infalling clumps are included and $Q(A)$
increases. $Q(A)$ then varies slowly, gradually falling as an increasing
number of randomly distributed noise peaks enter into the analysis.  Fig.~\ref{fig:dipEdS}b
shows $Q(A)$ for two clusters from the open $\Omega_{0}=0.2$
cosmology. $Q(A)$ is fairly stable over a wide range of area, never rising
above a value of $0.1$. This behaviour reflects the smooth and nearly
spherical mass distribution in these clusters. 

We now examine the distribution of $Q(A)$, at a fixed area $A$, for our
sample of clusters in each of the 3 cosmologies.  Figure~3 shows
cumulative distributions of $Q(A)$ obtained from the 24 clusters in each model
(see Section~\ref{sec:methods}). The three panels are for $A=3.8$
arcmin$^2$, $6.3$ arcmin$^2$, and $12.6$ arcmin$^2$. The two solid lines
are for $\Omega_0=1$, the two dotted lines for the open $\Omega_0=0.2$
model, and the two dashed lines for the flat $\Omega_0=0.2$ cosmology. 
For each pair, the line marked by triangles, usually with the smaller
quadrupole value, corresponds to the original, smoothed clusters. The
unmarked line corresponds to the reconstructed clusters. Examining the
original clusters first, it is apparent that an open $\Omega_0=0.2$
universe forms clusters with the smallest quadrupoles, followed by the
flat $\Omega_0=0.2$ model and finally the $\Omega_0=1$ case. After
reconstruction, the quadrupole value tends to increase, at least for the
$\Omega_0=1$ and the flat $\Omega_0=0.2$ clusters. This is primarily due
to the inclusion of outlying noisy pixels in the analysis.

Figure~3 shows that even with the imperfections of the weak lensing
reconstructions, there remains a strong and measurable difference in the
expected distribution of cluster shapes in high and low $\Omega_0$
cosmologies. The dependence of these distributions on $\lambda_0$ is very
weak and so this statistic cannot be used to constrain the cosmological
constant. This does mean, however, that $\Omega_0$ can be constrained
independently of the unknown value of $\lambda_0$. For example, for
$A=3.8$~arcmin$^2$ we see that $75\%$ of clusters in an $\Omega_0=1$
universe  have a $Q(A)$ value in excess of $0.1$, whereas less than $10\%$
of clusters in the two $\Omega_0=0.2$ universes are so aspherical.
Alternatively, the median value of $Q(A)$ is approximately $0.2$ for both
$A=3.8$ and~$6.3$~arcmin$^2$ in the  $\Omega_0=1$ universe, but less than
$0.06$ for both the $\Omega_0=0.2$ universes. These large differences 
suggest that weak lensing observations of a small number of clusters,
approximately five or so, can distinguish between these two values of
$\Omega_0$. A larger cluster sample is required to place finer constraints
on $\Omega_0$.

We stress that the distributions of $Q(A)$ plotted in Figure~3 are typical
of those expected from KS mass reconstructions using weak lensing data
obtained in realistic observing conditions. However, they are not intended
to be definitive descriptions of $Q(A)$ for these cosmologies. Clearly,
the exact form of the distributions will depend on the observing conditions,
the seeing and limiting magnitude, as well as on the redshifts of the
clusters. In practice, simulations that mimic specific observational datasets
will need to be performed for this test of $\Omega_0$ to be reliably applied. 

\section{Discussion and conclusions}

\label{sec:end}

We have shown how global properties of the dark matter distribution in
clusters, as revealed by weak gravitational lensing analyses, can be used
to measure the cosmological density parameter, $\Omega_0$. Our approach
complements and extends earlier work by \citeN{evr-94} who first
applied in practice the well established correlation between the
morphology of clusters and the underlying cosmology to estimate the value
of $\Omega_0$. Their analysis exploited the fact that the distribution of
hot X-ray emitting gas in clusters is expected to reflect the structure of the
underlying mass distribution. By probing this distribution directly,
gravitational lensing bypasses the need to assume that the morphology of
the gas faithfully mimics the morphology of the cluster mass. Possible
(although unlikely) non-gravitational processes that may affect this
correspondence are therefore not a concern, nor is there a worry that the
test may be affected by contaminating X-ray signals from, for example, AGNs.
A second advantage of the lensing approach is that it 
allows the matter distribution to be probed at larger cluster radii than
the centrally concentrated X-ray emission, that is, at radii where 
the distinction between different cosmological models is particularly 
strong. 

The major disadvantage of the lensing method is its sensitivity to
projection effects. The observed galaxy shear pattern is a 
function of the product of the projected surface density along the 
line-of-sight
and the effective cross-section for lensing (which depends on distance
through equation~\ref{eq:scrit}). Mass clumps at large distances from the cluster
contribute to the lensing signal and, in principle, to the measured
quadrupole. In practice, this is unlikely to be a strong effect,
particularly when the analysis is restricted to relatively small areas
around the cluster centre.  Nevertheless the size of this contamination
needs to be assessed using different simulations from those discussed
here. X-ray analyses are much less sensitive to projection effects and so
the two approaches are complementary and should be used in combination.

In some respects, our detailed results must be regarded as preliminary. The
simulations by \citeANP{evr-94} which we have analyzed have a number of
limitations. Firstly, the initial conditions in all the cosmological models
were laid down with the power spectrum appropriate to an $\Omega_{0}=1$ CDM
universe. This inconsistency between the power spectrum and the
cosmology was further exacerbated in our case by the
need to rescale the simulations in order to ensure a lensing signal with
comparable signal-to-noise in all cosmologies. Secondly, the baryon
fraction in all the simulations was fixed to be 10\% of the critical
density so that in the $\Omega_{0}=0.2$ models, the gas contributes half of the
total gravitating mass. Finally, the clusters we have analyzed were chosen
to correspond to high peaks in the density field, but they do not represent
a proper statistical sample. Although we argue that all these
approximations are unlikely to affect our results significantly, it is
clearly desirable to repeat our analysis with purpose-built simulations.

The main result of our work is that the distribution of a low-order
statistic sensitive to global deviations from spherical symmetry in the
mass distribution of clusters depends strongly on $\Omega_0$. The
distribution of this quadrupole statistic can be robustly recovered
from the shear pattern of galaxies weakly lensed by the cluster
gravitational potential. We have shown that under standard observing
conditions, lensing data for a handful of clusters should distinguish
between cosmological models with $\Omega_0=1$ and $\Omega_0=0.2$.

\section*{Acknowledgements}

We would like to thank Gus Evrard for very
kindly supplying us with his cluster simulations. We would
also like to thank Chris Metzler for many useful discussions.
SMC acknowledges the support of a PPARC Advanced Fellowship.

\bibliography{lensing}
\bibliographystyle{mnras}

\bsp

\end{document}